\documentclass[twocolumn,10pt,cleanfoot]{cav2012}

\usepackage{graphicx}
\usepackage{amsmath}
\confshortname{CAV2012}
\conffullname{the 8th International Symposium on Cavitation}
\confdate{14-16}
\confmonth{August}
\confyear{2012}
\confcity{Singapore}
\confcountry{Singapore}

\papernum{\#\#}

\title{Environment-induced heating in sonoluminescence experiments}

\author{Almut Beige
\affiliation{School of Physics and Astronomy, \\
University of Leeds, Leeds, LS2 9JT, \\
United Kingdom}}

\author{ Antonio Capolupo
\affiliation{Dipartimento di Ingegneria Industriale, \\
Universita' di Salerno, 84100 Salerno, \\
Italy}}

\author{Andreas Kurcz
\affiliation{School of Physics and Astronomy, \\ 
University of Leeds, Leeds, LS2 9JT, \\
United Kingdom}}

\begin{document}

\maketitle

\begin{abstract}
{\it This paper discusses a quantum optical heating mechanism which might play an important role in sonoluminescence experiments. We suggest that this mechanism occurs during the final stages of the bubble collapse phase and accompanies the thermodynamic heating due to the compression of the bubble. More concretely, it is shown that a weak but highly inhomogeneous electric field, as it occurs naturally during rapid bubble deformations, can increase the temperature of strongly confined particles by several orders of magnitude on a nanosecond time scale [A. Kurcz {\em et al.}, New J. Phys. {\bf 11}, 053001 (2009)]. Our model suggests that it is possible to control sonoluminescence experiments in ionic liquids with the help of appropriately detuned laser fields.}
\end{abstract}

\section*{INTRODUCTION}
Sonoluminescence is the intriguing phenomenon of strong light flashes from tiny bubbles in a liquid \cite{Brenner}. The bubbles are driven by an ultrasonic wave and need to be filled with atomic species. Approximating the emitted light by blackbody radiation indicates very high temperatures. Although sonoluminescence has been studied extensively, a complete understanding of the mechanisms which concentrate the energy of acoustic vibrations seems still missing \cite{Flannigan}. Moreover, the actual role of atomic species inside the cavitating bubble seems to remain controversial. The main purpose of this paper is to emphasise that quantum optical heating mechanisms of trapped particles via highly inhomogenous electric fields can play an important role in creating very high temperatures in sonoluminescence experiments \cite{sono}. Moreover, we suggest that it is possible to increase the temperature inside the cavitating bubble even further with the help of appropriately detuned laser fields.

To do so, we emphasise that energy-absorbing measurements on one component of a composite quantum system can have a huge effect on the energy of the total  system \cite{ec2}. In addition, we point out close analogies between ion trap and sonoluminescence experiments. For example, it is well known that it is possible to cool ions in ion trap experiments with red-detuned laser fields to nanokelvin temperatures \cite{cooling,cooling2,JMO}. Blue detuned laser fields heat ions in ion trap experiments very rapidly and result therefore in their immediate rejection from the trap. Taking a more general point of view than in our earlier work \cite{sono}, this paper tries to provide more insight into possible quantum optical heating effects in sonoluminescence experiments.

In the next section, we consider composite quantum systems with repeated energy-absorbing measurements on one systems component. A simple example of such a system is a well localised atom (or ion) in free space which interacts with the surrounding free radiation field \cite{Hegerfeldt,Dalibard,Carmichael,Adam}. Another example is a trapped atom (or ion) with quantised motional states in a free radiation field \cite{cooling,cooling2}. In these two systems, repeated energy-absorbing measurements on one system component -- the free radiation field -- occur naturally in the presence of a photon-absorbing environment, like the walls of the laboratory. When the atom (or ion) is driven by external laser fields, these measurements manifest themselves in the spontaneous emission of photons. Quantum optical models based on the assumption of environment-induced photon-absorbing measurements are in general in very good agreement with experiments \cite{one,two}. 

In the third section of this paper, we emphasise common features between ion trap and sonoluminescence experiments. More concretely, we assume that the particles inside the cavitating bubble are so strongly confined that their motion becomes quantised. In addition, we assume that a weak but highly inhomogenous electric field establishes a relatively strong coupling between the electronic states and the motional states of the trapped particles. Such a field occurs naturally during rapid bubble deformations. For example, it can be generated via the formation of a plasma inside the cavitating bubble \cite{plasma,plasma2}. The effect of an inhomogenous electric field is similar to the effect of laser driving in ion trap experiments. Using a quantum optical master equation \cite{sono}, we predict a rapid exponential growth of the mean number of phonons per particle, ie.~heating. In the case of rapidly changing system parameters, numerical simulations predict an even faster temperature growth \cite{sono}. 

Finally, we point out that the above close analogy between ion trap and sonoluminescence experiments suggests that it is possible to significantly increase the temperature inside a cavitating bubble via the application of blue-detuned laser light. Recent measurements of the spectrum of the emitted light in sonoluminescence experiments in ionic liquids show relatively sharp emission lines in the optical regime \cite{Flannigan,species}. These verify the availability of atomic transitions which could be driven by laser light during the final stages of the bubble collapse phase. Choosing the detuning of the applied laser fields appropriately should allow us to control the temperature of the particles inside a cavitating bubble analogously to the manipulation of the temperature of ions in ion trap experiments. Since the corresponding heating rates can be very high \cite{prep}, we expect that it is possible to significantly increase the temperature of the particles inside a cavitating bubble with blue-detuned laser fields. 

\section*{HEATING QUANTUM SYSTEMS BY MEASUREMENTS}
In quantum systems with unitary evolutions, the total energy is a conserved quantity, since the corresponding
Hamiltonian $H$ always commutes with itself $([H,H]=0)$. However, this argument does not apply to open quantum systems. For example, the spontaneous emission of a photon is always related to the loss of energy from its source \cite{Adam}. In this section, we discuss a mechanism which can be used to change the energy of a composite quantum system. To do so, we consider a composite quantum system with two interacting components $A$ and $B$. In addition, we assume that system $B$ experiences a sequence of energy-absorbing measurements.  

Let us first have a closer look at a single energy-absorbing measurement on a quantum system. For simplicity, we  consider a system with discrete energy eigenvalues $E_i$ and denote its energy eigenstates by $|\lambda_i \rangle$ with $|\lambda_0 \rangle$ being the ground state. Suppose the system is initially in $|\psi \rangle$. Then the probability of measuring $E_i$ equals 
\begin{eqnarray}
p_i = |\langle \lambda_i | \psi \rangle |^2 \, . 
\end{eqnarray}
In case of an energy-absorbing measurement, the system is immediately after the measurement in its lowest energy eigenstate $|\lambda_0 \rangle$. This means, repeating the measurement, the lowest energy value $E_0$ is obtained with probability one.

We now have a closer look at a composite quantum system with two interacting subsystems $A$ and $B$. The Hamiltonian $H$ of such a system can always be written as 
\begin{eqnarray} \label{Hsys}
H &=& H_A + H_B + H_{\rm int} \, ,
\end{eqnarray}
where $H_A$ accounts for the energy of subsystem $A$, $H_B$ accounts for the energy of subsystem $B$, while $H_{\rm int}$ denotes the interaction between $A$ and $B$. Suppose, an energy-absorbing measurement is performed on subsystem $B$. The possible outcomes of this measurement are the energy eigenvalues $E_i$ and the energy eigenstates $|\lambda_i \rangle_B$ of the Hamiltonian $H_B$. If the total system is initially in a state $|\psi \rangle_{AB}$, the measurement initiates a transformation of the form
\begin{eqnarray} \label{not}
|\psi \rangle_{AB} &\longrightarrow & |\psi \rangle_A \otimes |\lambda_0 \rangle_B \, .
\end{eqnarray}
Notice that the state on the right hand side of this equation is in general not an energy eigenstate of the total Hamiltonian $H$ in Eq.~(\ref{Hsys}). This means, it subsequently evolves in time. During this time evolution, the energy of the composite quantum system with respect to its Hamiltonian $H$ in Eq.~(\ref{Hsys}) does not change but energy can leak from one subsystem into another. This means, the energy of the subsystem $A$ and subsystem $B$ with respect to $H_A$ and $H_B$, respectively, can change significantly.

More concretely, we now consider a short time interval $(t,t+\Delta t)$ and assume that the total system is initially in the state $ |\psi \rangle_A \otimes |\lambda_0 \rangle_B$ which forms the right hand side of Eq.~(\ref{not}). Then it evolves such that
\begin{eqnarray} \label{not2}
|\psi \rangle_A |\lambda_0 \rangle_B 
&\longrightarrow& U(t,t+\Delta t) \, |\psi \rangle_A |\lambda_0 \rangle_B \, .
\end{eqnarray}
If a second energy-absorbing measurement is performed on subsystem $B$ at $t+\Delta t$, the probability to obtain an amount of energy $E_i$ larger than $E_0$ equals 
\begin{eqnarray}
p_{i>0} &=& \sum_{i>0} \| \, _B \langle \lambda_i| U(t+\Delta t,t) |\psi \rangle_A |\lambda_0 \rangle_B \, \|^2 \, .
\end{eqnarray}
This probability is in general larger than zero. When forgetting about the presence of the interaction $H_{\rm int}$ and the effect of a measurement on single quantum system, it might look as if energy appears from nowhere \cite{ec2}. 

To illustrate this effect even more clearly, assume for example that the composite quantum system is initially in its energy ground state $|\lambda_0 \rangle_{AB}$ with respect to $H$ in Eq.~(\ref{Hsys}). During an energy-absorbing measurement on subsystem $B$, this state changes such that
\begin{eqnarray} \label{not3}
|\lambda_0 \rangle_{AB} 
&\longrightarrow& |\lambda_0 \rangle_{\rm BB} \langle \lambda_0 | \lambda_0 \rangle_{AB} / \| \cdot \| \, .
\end{eqnarray}
The state on the right hand side of this equation is in general not the energy groundstate of the total system. When this applies, it must be a superposition of several energy eigenstates $| \lambda_i \rangle_{AB}$ of $H$ with $i \ge 0$. The overall effect of the energy-absorbing measurement on subsystem $B$ on the state of the total system is hence in general a transition of the form 
\begin{eqnarray} \label{not4}
|\lambda_0 \rangle_{AB} 
&\longrightarrow& \sum_{i \ge 0} \xi_i \, | \lambda_i \rangle_{AB} \, ,
\end{eqnarray}
where the $\xi_i$ are complex coefficients with $\sum_{i \ge 0} | \xi_i |^2 = 1$. This means, the energy-absorbing measurement on subsystem $B$ increased the expectation value of the total energy of the system. It now equals
\begin{eqnarray} \label{E}
\langle H \rangle &=& \sum_{i \ge 0} | \xi_i |^2 \, E_i \, .
\end{eqnarray}
Here, the $E_i$ denote the energy eigenvalues of $H$. The subsequent evolution of the system redistributes this energy within the system such that some of it can be extracted in a subsequent energy-absorbing measurement on subsystem $B$. Of course, this process is in general also accompanied by a significant change of the energy of subsystem $A$ with respect to $H_A$. Heating and cooling of subsystem $A$ can occur. 

\section*{HEATING IN SONOLUMINESCENCE EXPERIMENTS}
The previous section illustrates that energy-absorbing measurements on one component of a composite quantum system can change the energy of its subsystems significantly. The aim of this paper is to point out that such an effect could play an important role in sonoluminescence experiments. To do so, we identify subsystem $A$ with a strongly confined particle which is typical for the many particles inside the cavitating bubble. This means, subsystem $A$ can be a nobel gas atom or an ion from an ionic liquid \cite{Flannigan,species}. As in Ref.~\cite{sono}, we assume that the particle experiences a strong enough trapping potential such that its motion becomes quantised. Moreover, we identify subsystem $B$ with the free radiation field which surrounds a transparent bubble. This field couples to the electronic states of the trapped particle, thereby resulting in a Hamiltonian of the same form as $H$ in Eq.~(\ref{Hsys}). In addition, there are energy-absorbing measurements on subsystem $B$, since the free radiation field interacts with a photon-absorbing environment \cite{Hegerfeldt,Adam}. 

Following the ideas of the previous section, one might expect that a trapped atom (or ion) in a free radiation field emits photons with a constant rate, even in the absence of external driving \cite{Adam}. This is not the case, since the ground state of the Hamiltonian $H$ of a trapped particle inside a free radiation field in the vacuum state is also the ground state of the free Hamiltonian $H_A + H_B$. This means, $|\lambda_0 \rangle_{AB} = |\lambda_0 \rangle_A \otimes |\lambda_0 \rangle_B $ in this case. However, this no longer applies in the presence of a weak but highly inhomogenous electric field inside the cavitating bubble. Such a field might occur naturally during rapid bubble deformations due to one or more of the following mechanisms (with the last one being the most relevant one):
\begin{enumerate}
\item The water dipole molecules are pushed out of their equilibrium position when the bubble radius shrinks suddenly. It takes some time before they can re-arrange themselves again in a ``neutral" position.  
\item The sudden vapourisation of water molecules during the bubble collapse phase causes particles with a non-zero dipole moment to move rapidly through the bubble.
\item Once the bubble reaches a certain temperature, a plasma is formed which is accompanied by highly inhomogenous electric fields and rapidly moving charged particle (ions and electrons) within the bubble \cite{plasma,plasma2}. 
\end{enumerate}
This means, at the final state of the collapse phase, a trapped particle inside a cavitating bubble becomes an example of the composite quantum system which we described in the previous section. As we shall see below, a photon-absorbing environment indeed distributes energy into its motional degrees of freedom. When analysing the dynamics of a single trapped particle using a quantum optical master equation \cite{cooling,cooling2}, we observe an exponential increase of the mean phonon number per particle on a nanosecond time scale \cite{sono}. Rapidly changing system parameters enhance this heating process even further \cite{sono}. The remainder of this section reviews the main results of our earlier work.

\subsection*{Theoretical model}

As mentioned already above, in this subsection, we consider a single trapped atom (or ion) which is typical for the many particles inside the cavitating bubble. When excited, the atom is able to spontaneously emit a photon which indicates a significant coupling to the surrounding free radiation field. When the bubble reaches its minimum radius, the motion of the atom becomes so strongly confined that it becomes quantised. In addition, we assume the presence of a relatively weak but highly inhomogenous electric field ${\bf E}$ at the position ${\bf r}$ of the atom which is of the form
\begin{eqnarray} \label{E}
{\bf E} ({\bf r}) &=& \sum_k {\bf E}_k \, {\rm e}^{{\rm i} k \hat {\bf k} \cdot {\bf r}} + {\rm c.c.} 
\end{eqnarray}  
The Hamiltonian $H$ of this system in the Schr\"odinger picture is of the same form as $H$ in Eq.~(\ref{Hsys}). More concretely, we have \cite{sono,Hegerfeldt,Adam}
\begin{eqnarray} \label{Hs}
H_A &=& \hbar \omega_0 \, \sigma^+ \sigma^- + \hbar \nu \, b^\dagger b + e \, {\bf D} \cdot {\bf E}({\bf r}) \, , \nonumber \\
H_B &=& \sum_{{\bf k}\lambda} \hbar \omega_k \, a_{{\bf k}\lambda}^\dagger a_{{\bf k}\lambda} \, , \nonumber \\
H_{\rm int} &=& \sum_{{\bf k}\lambda} \hbar g_{{\bf k}\lambda} \big( \sigma^+ a_{{\bf k}\lambda} + \sigma^- a_{{\bf k}\lambda}^\dagger \big) \, .
\end{eqnarray}
Here, $\sigma^- = |0 \rangle \langle 1|$ and $\sigma^+ = |1 \rangle \langle 0|$ are the lowering and the raising operator for a two-level atom with $|0 \rangle$ being its ground state and $|1 \rangle$ being its excited state with energy $\hbar \omega_0$. The operators $b$ and $b^\dagger$ are the bosonic annihilation and creation operator for a single phonon with frequency $\nu$. Moreover, $a_{{\bf k}\lambda}$ and $a_{{\bf k}\lambda}^\dagger$ are the bosonic annihilation and creation operator, respectively, for a photon in the free radiation field with wave vector ${\bf k}$, frequency $\omega_k$, and polarisation $\lambda$, while the $g_{{\bf k}\lambda}$ describe the coupling of the photon modes $({\bf k},\lambda)$ to the atom. 

Let us now have a closer look at the effect of ${\bf E} ({\bf r})$. In Eq.~(\ref{Hs}), $e$ is the charge of a single electron and ${\bf D}$ is the atomic dipole moment operator ${\bf D} =  {\bf D}_{01} \, \sigma^{-} + {\rm H.c.}$ with ${\bf D}_{01}$ being a (real) three-dimensional vector. When replacing the atomic position ${\bf r}$ with respect to its equilibrium position by the atomic position operator (which equals $b + b^\dagger$ up to a constant) and applying the usual Lamb-Dicke approximation \cite{LambDicke} for a particle which is well localised at ${\bf r}=0$, we find that
\begin{eqnarray}
{\rm e}^{{\rm i} k \hat {\bf k} \cdot {\bf r}} &=& 1 + {\rm i} k \Delta x \big(b + b^\dagger \big) 
\end{eqnarray}  
to a very good approximation. The space $\Delta x$ occupied by the particle is to a very good approximation given by \footnote{For simplicity, we consider only a single phonon frequency $\nu$, ie.~we assume a harmonic trapping potential for the particle, although a continuous range of frequencies should be taken into account. A more realistic model should be based on a square well or a van der Waals trapping potential. However, such a model would be more complicated but yields nevertheless  similar results.}
\begin{eqnarray} \label{x}
\Delta x &=& \left( {\hbar \over 2 M \nu} \right)^{1/2} \, , 
\end{eqnarray}
where $M$ denotes the mass of the respective atom. Combining this expression with the above equations, we find that the interaction between the atomic states and an inhomogenous electric field can be written as \cite{sono}
\begin{eqnarray} \label{Hint}
e \, {\bf D} \cdot {\bf E}({\bf r}) &=& \hbar \Omega \, ( \sigma ^- + \sigma^+) + \hbar \Lambda  \, (b  + b ^\dagger) ( \sigma ^- + \sigma^+)
\end{eqnarray}
with the (real and positive) coupling constants 
\begin{eqnarray} \label{shrink}
\Omega &\equiv & (2 e / \hbar) \sum _k {\bf D}_{01} \cdot {\rm{Re}} \left. \left ( {\bf E}_k {\rm e}^{{\rm{i}} k \hat {\bf k} \cdot \bf{r}} \right ) \right|_{{\bf r}=0} \, , \nonumber \\
\Lambda &\equiv & - (2 e \Delta x / \hbar) \sum _k k  \, {\bf D}_{01} \cdot {\rm{Im}} \left. \left ( {\bf E}_k {\rm e}^{{\rm{i}} k \hat {\bf k} \cdot {\bf r}} \right ) \right|_{{\bf r}=0} \, .
\end{eqnarray} 
A closer look at Eq.~(\ref{shrink}) shows that $\Lambda$ is proportional to the gradient of $\Omega$ in the direction of the quantised motion of the atom. It can be written as
\begin{eqnarray} \label{shrink2}
\Lambda &=& \Delta x \, \hat{\bf k} \cdot \nabla \Omega ({\bf r}) \, . 
\end{eqnarray}
This equation shows that a strong atom-phonon coupling does not require a strong but a highly inhomogeneous electric field inside the bubble. 

Spontaneous photon emission is in the following taken into account via a master equation approach based on the Hamiltonian $H$ in Eq.~(\ref{Hs}) and on the assumption of rapidly repeated photon-absorbing measurements on a coarse grained time scale \cite{Hegerfeldt,Dalibard,Carmichael,Adam}. Using this approach, we find that the atomic density matrix $\rho_A$ for the electronic and the vibrational states of the trapped particle inside the cavitating bubble evolves according to the differential equation
\begin{eqnarray} \label{master2}
\dot \rho_A = - \frac{{\rm i}}{\hbar} \big[ H _A ,  \rho_A \big] + {1 \over 2} \Gamma \big[ 2 \sigma ^- \rho_A \sigma ^+ - \sigma ^+ \sigma ^- \rho_A - \rho_A \sigma ^+ \sigma ^- \big] \, . \nonumber \\
\end{eqnarray}
Here $\Gamma$ denotes the spontaneous photon emission rate of the excited electronic state $|1 \rangle$. An analogous master equation approach is routinely used in the literature to model laser driving in ion trap experiments \cite{cooling,cooling2}.

\subsection*{Time evolution of the mean phonon number}

In Ref.~\cite{sono}, we used Eq.~(\ref{master2}) to analyse the time evolution of the mean number $m$ of phonons per trapped particle,
\begin{eqnarray}
m &\equiv & \langle b^\dagger b \rangle \, ,
\end{eqnarray}
with the help of a closed set of rate equations, ie.~a closed set of linear differential equations for expectation values. In the strong atom-phonon coupling regime, where
\begin{eqnarray} \label{par} 
\Lambda \, \gg \, \Omega  ~~& {\rm and} &~~ 4 \Lambda^2 \, > \, \nu \omega_0 \, ,
\end{eqnarray} 
and for time-independent parameters, we find that $m(t)$ equals 
\begin{eqnarray} \label{fullsolution2}
m(t) &=& \left[ 1 + {8 \Lambda^4 \over (\lambda \omega _0)^2} \, \sinh^2 ( \lambda t ) \right] \, m(0)
\end{eqnarray}
to a very good approximation, when we define
\begin{eqnarray} \label{lambda}
\lambda &\equiv & \nu \left( {4 \Lambda^2 \over \nu \omega _0} - 1 \right)^{1/2} \, .
\end{eqnarray}
Eq.~(\ref{fullsolution2}) describes an approximately exponential heating process with $2 \lambda $ being the corresponding heating rate. This process takes place on a nanosecond time scale, for example, when $\nu = 10\,$MHz, $\Omega = 10^{6} \, $Hz, $\Lambda = 10^{12} \, $Hz, $\Gamma = 10^{13} \, $Hz, and $\omega_0 = 10^{15} \, $Hz (cf.~Fig.~3(c) in Ref.~\cite{sono}). When considering time-dependent parameters, the above mentioned rate equations can be used to reveal a significant speed up of the heating process (cf.~Fig.~5 in Ref.~\cite{sono}).

\subsection*{Atomic level scheme}

\begin{figure}[t]
\centering
  \includegraphics[width=0.4\textwidth]{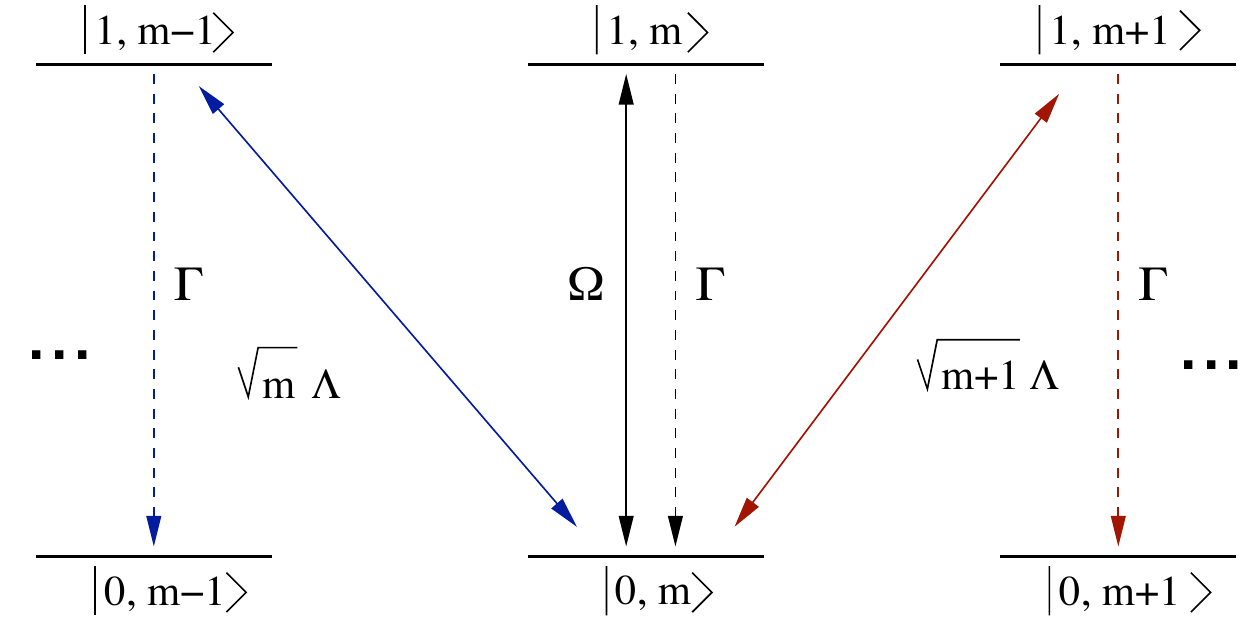}
  \caption{Level configuration of a single atom-phonon system indicating the immediately relevant transitions, if the atom is initially in its ground state $|0 \rangle$ and possesses exactly $m$ phonons. As in Eq.~(\ref{Hint}), $\Omega$ denotes the coupling constant for phonon conserving transitions to the excited atomic state $|1 \rangle$, while $\Lambda $ is due to the electric field gradient inside the bubble and establishes a coupling between the electronic and the motional states of the atom. Moreover, $\Gamma$ is the spontaneous photon decay rate of level 1.} \label{fig2}
\end{figure}

An alternative perspective on the origin of the above described heating process is illustrated in Fig.~\ref{fig2}. This figure shows the main transitions involved in the heating process, if the trapped particle is initially in its electronic ground state $|0 \rangle$ and there are exactly $m$ phonons in its vibrational mode. In this case, the presence of a highly inhomogenous electric field can result in the excitation of the atom which is accompanied either by the creation or the annihilation of a phonon (cf.~Eq.~(\ref{Hint})). Taking into account that phonons are bosonic particles and that the annihilation operator $b$ obeys the bosonic commutator relation 
\begin{eqnarray} \label{commi}
[b,b^\dagger] &=& 1 \, ,
\end{eqnarray}
one can show that 
\begin{eqnarray} \label{commi2}
b \, |m \rangle &=& \sqrt{m} \, |m -1 \rangle \, , \nonumber \\
b^\dagger \, |m \rangle &=& \sqrt{m+1} \, |m +1 \rangle  
\end{eqnarray}
for all phonon number states $|m \rangle$. This means, the creation of a phonon is always slightly more likely than the annihilation of a phonon. As a consequence, the master equation in Eq.~(\ref{master2}) does not possess a stationary state solution. The mean phonon number $m$ increases until the inhomogenous electric field inside the bubble becomes too small and the coupling constant $\Lambda$ no longer obeys the conditions in Eq.~(\ref{par}).

\section*{LASER HEATING OF SONOLUMINESCING BUBBLES}
The previous subsection reveals a close analogy between ion trap and sonoluminescence experiments. For example, in both kinds of experiments, particles experience very strong trapping potentials. One way of obtaining a rough estimate for the relevant phonon frequency $\nu$ in a cavitating bubble is to notice that the volume of the bubble equals $V_{\rm bubble} = {4 \over 3} \pi R^3$, if $R_{\rm bubble} $ is its radius. This means, on average, each atom occupies a volume of $V_{\rm atom} = V_{\rm bubble}/N$, where $N$ is the number of atoms inside the bubble. The radius $R_{\rm atom} = R_{\rm bubble} /N^{1/3}$ which corresponds to this volume can be used as an estimate for $\Delta x$ in Eq.~(\ref{x}). For $N=10^5$ and $R= 500\,$nm, we obtain a $\Delta x$ of about $10 \,$nm. The Argon atom mass of $M=6.63 \cdot 10^{-26} \,$kg hence suggests typical phonon frequencies $\nu$ for Argon ions of about $10\,$MHz.\footnote{If we assume instead that $N=10^{10}$, then $\Delta x$ reduces to $2.3$ Angstrom, which corresponds to a value close to the van der Waals volume for Argon and an even higher phonon frequency $\nu$.} Due to the presence of other particles inside the bubble, this number is probably only a lower bound. For comparison, let us mention that the phonon frequency $\nu$ of ion trap experiments is usually of the order of $1 \,$MHz \cite{cooling,cooling2}. 

Moreover, recent measurements of the spectrum of the emitted light in sonoluminescence experiments in ionic liquids show relatively sharp emission lines in the optical regime \cite{Flannigan,species}. These verify the availability of atomic transitions which could be driven by laser light. The fact that there are strongly confined particles inside a cavitating bubble therefore suggests that it is possible to control the temperature of a cavitating bubble via the application of appropriately detuned laser fields. Like an inhomogeneous electric field, laser fields are able to establish a strong coupling between the electronic and the motional states of trapped particles. Combined with a non-zero spontaneous photon emission rate, they are able to promote the creation of phonons at a much higher rate than the annihilation of phonons. This means, depending on the respective laser detunings, they should be able to cool and to heat the particles inside a cavitating bubble on very short time scales \cite{prep}.

More concretely, when applying a laser field, we essentially replace the electric field ${\bf E} ({\bf r}) $ in Eq.~(\ref{E}) by a time-dependent electric field of the form
\begin{eqnarray}
{\bf E} ({\bf r}) &=& {\bf E}_0 \, {\rm e}^{{\rm i} ({\bf k} \cdot {\bf r} + \omega_{\rm L} t)} + {\rm c.c.} 
\end{eqnarray}
Here $\omega_{\rm L}$ is the frequency of the applied laser field. To remove the time-dependence from this Hamiltonian, we change into an interaction picture with respect to $H_0 = \hbar \omega_{\rm L} \, \sigma^+ \sigma^-$. Doing so and applying again the Lamb-Dicke approximation \cite{LambDicke} for well localised particles, we now obtain the atomic Hamiltonian 
\begin{eqnarray} \label{Hs2}
H_A &=& \hbar (\omega_0 - \omega_{\rm L}) \, \sigma^+ \sigma^- + \hbar \nu \, b^\dagger b + e \, {\bf D} \cdot {\bf E}({\bf r}) \, ,
\end{eqnarray}
where the last term can now be written as
\begin{eqnarray} \label{Hint2}
e \, {\bf D} \cdot {\bf E}({\bf r}) &=& \hbar \Omega \, ( \sigma ^- + \sigma^+) + \eta \hbar \Omega  \, (b  + b ^\dagger) ( \sigma ^- + \sigma^+) \, . ~~
\end{eqnarray}
The parameter $\Omega \equiv 2e{\bf D}_{01} \cdot {\bf E}_0^*/\hbar$ in this equation is the laser Rabi frequency, while $\eta$ is the Lamb-Dicke parameter and usually about one order of magnitude smaller than one. A comparison of Eqs.~(\ref{Hs2}) and (\ref{Hint2}) with Eqs.~(\ref{Hs}) and (\ref{Hint}) shows that there is indeed a close analogy between the situation described here and the situation analysed in the previous section. When analysing the time evolution of the mean phonon number $m$, we again find an exponential evolution \cite{cooling,cooling2}. 

\begin{figure}[t]
\centering
  \includegraphics[width=0.40\textwidth]{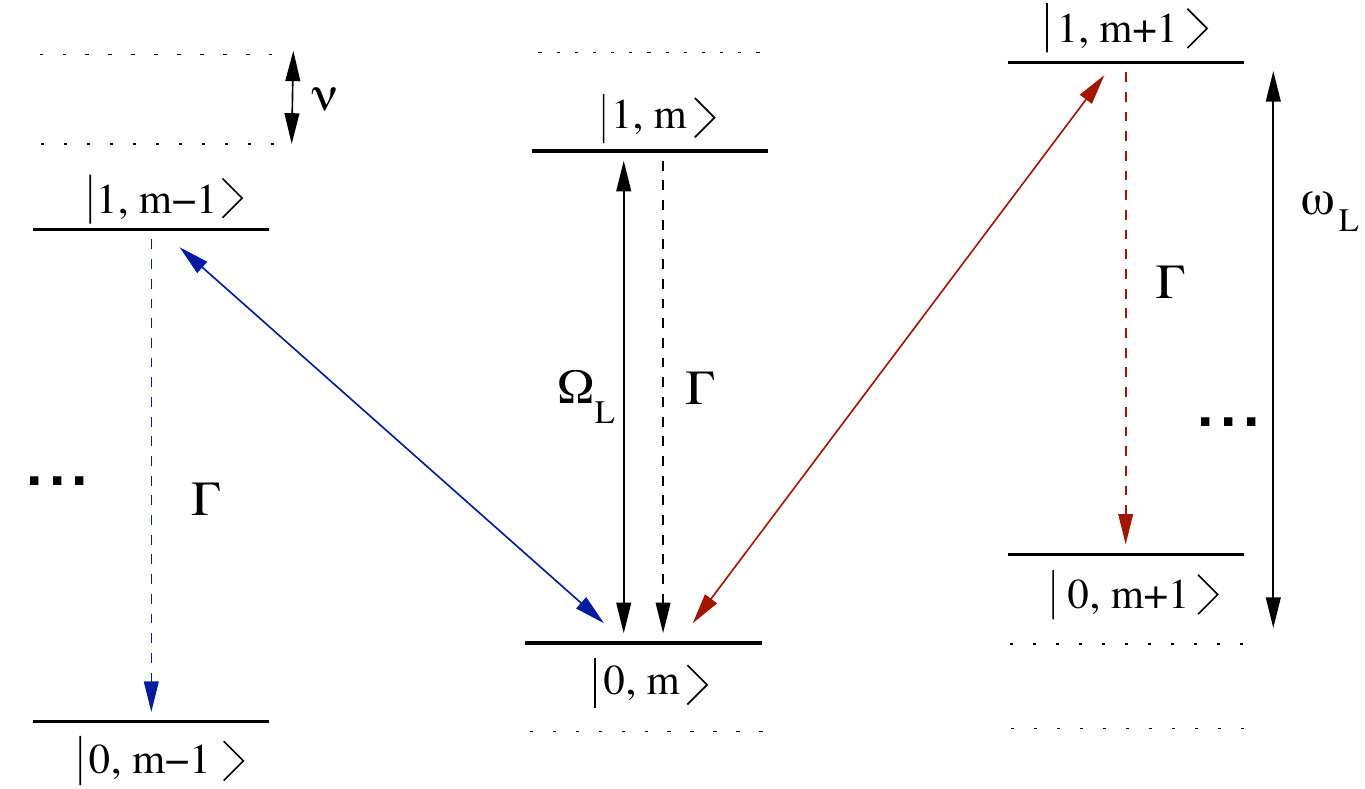}
  \caption{Illustration of a mechanism to enhance the energy concentration in sonolumiscence experiments via laser heating. Shown are the relevant transitions in a single atom-phonon system, when the atom is in its ground state $| 0 \rangle$ and possesses exactly $m$ phonons, while a laser field excites the blue sideband resonantly.}\label{fig7}
\end{figure}

However, notice that the laser frequency $\omega_{\rm L}$ can be chosen such that $|\omega_0 - \omega_{\rm L}|$ is many orders of magnitude smaller than $\omega_0$. This means, the desired atomic transitions can become resonant with the systems dynamics. In contrast to this, the atomic transitions induced by a highly inhomogenous electric field which we described in the previous section are detuned by the atomic transition frequency $\omega_0$. This is the reason why the effective heating rate $\lambda$ in Eq.~(\ref{lambda}) scales essentially as $1/\sqrt{\omega_0}$. For laser-induced heating processes, like the one illustrated in Fig.~\ref{fig7}, this is no longer the case. Laser-induced heating can hence be much more effective than quantum optical heating due to the presence of a highly inhomogenous electric field. 

Experience with ion trap experiments \cite{cooling,cooling2} suggests that maximum heating rates occur, when the laser field excites the blue sideband resonantly. In other words, one should choose the laser frequency $\omega_{\rm L}$ close to $\omega_0 + \nu$ such that transitions which excite the atom and create a phonon become resonant and are hence the most likely transitions to occur. The dynamics of a trapped atom with laser driving on the blue sideband is illustrated in Fig.~\ref{fig7}. There we assume again that the atom is initially in its electronic ground state $|0 \rangle$ and that there are exactly $m$ phonons in its vibrational mode. Suppose a blue-detuned laser field excites the state $|1,m+1 \rangle$. When followed by the spontaneous emission of a photon, the state of the system becomes $|0, m+1 \rangle $. Overall, a phonon has been gained \cite{cooling,cooling2}. 

\section*{CONCLUSIONS}

This paper reviews a quantum optical heating mechanism \cite{sono} which might play an important role during the final stages of the bubble collapse phase in sonoluminescence experiments. It is shown that a weak but highly inhomogeneous electric field, as it might occur naturally during rapid bubble deformations, can increase the temperature of strongly confined particles very rapidly. The analysis of the proposed heating process in this paper and in our earlier paper (cf.~Ref.~\cite{sono}) is rather qualitative than quantitative, since it is hard to directly calculate typical values for the coupling constant $\Lambda$. The reason for this is that $\Lambda$ relates to the gradient of the inhomogenous field inside the cavitating bubble. Consequently, it is difficult to determine concrete values for the heating rate $\lambda$ in Eq.~(\ref{lambda}). However, since all relevant quantities are quantum optical in nature, we assume that the described heating process can increase the mean number of phonons per trapped particle by several order of magnitude, even on a nanosecond time scale. 

What supports our thesis of the presence of a quantum optical heating mechanism in sonoluminescence experiments is the importance of the presence of atomic species inside the cavitating bubble. High temperatures in sonoluminescence experiments are achieved, for example, when there is a high concentration of nobel gas atoms \cite{nobel}. Compared to normal atoms and ions, nobel gas atoms are unlikely to be involved in chemical reactions. However, the electronic transitions of the valence electrons of nobel gas atoms lie in the ultraviolet regime \cite{transitions}. This means, their $\omega_0$ is significantly larger than the typical transition frequencies $\omega_0$ of other particle species which lie in in the optical regime. For example, typical transition frequencies of Argon atoms are above $13 \,$eV, while transitions frequencies in the optical regime are below $1 \,$eV \cite{transitions,transitions2}. As a result, particles with transitions in the optical regime can result in higher heating rates $\lambda$ (cf.~Eq.~(\ref{lambda})) then the typical $\lambda$ of a nobel gas atom. Recently, it has been found that sonoluminescence experiments in ionic liquids can exhibit much higher photon emission rates than sonoluminescence experiments with nobel gas atoms \cite{liquid,liquid2}.

The presence of atomic transitions in the optical regime during the bubble collapse phase and the presence of trapping frequencies $\nu$ which are comparable to the trapping frequencies $\nu$ in ion trap experiments suggests that it is possible to control the temperature in sonoluminescence experiments, especially in ionic liquids, with the help of appropriately detuned laser fields. Maximum rates for the atomic heating process are expected when the applied laser field excite atomic transitions close to blue sideband \cite{sono}. These heating rates can be much larger than the heating rates $\lambda$ induced by a highly inhomogenous electric field, since the relevant atomic transitions are now resonant instead of being essentially detuned by the atomic transition frequency $\omega_0$.

\section*{ACKNOWLEDGMENTS}
The authors acknowledge stimulating discussions with E. Del Giudice and G. Vitiello. This work has been supported by the United Kingdom Research Council EPSRC.

\nocite*
\bibliographystyle{Cav2012}

\end{document}